# Band gap engineering by Bi intercalation of graphene on Ir(111)

*Jonas Warmuth[1], Albert Bruix[2], Matteo Michiardi[2], Torben Hänke[1], Marco Bianchi[2], Jens Wiebe[1], Roland Wiesendanger[1], Bjørk Hammer[2], Philip Hofmann[2], Alexander A. Khajetoorians[3,1,*]*

[1.] Department of Physics, Hamburg University, Hamburg, Germany

[2.] Interdisciplinary Nanoscience Center and Department of Physics and Astronomy, Aarhus University, Denmark

[3.] Institute of Molecules and Materials, Radboud University, Nijmegen, Netherlands

[*]Correspondence to: a.khajetoorians@science.ru.nl

We report on the structural and electronic properties of a single bismuth layer intercalated underneath a graphene layer grown on an Ir(111) single crystal. Scanning tunneling microscopy (STM) reveals a hexagonal surface structure and a dislocation network upon Bi intercalation, which we attribute to a $\sqrt{3} \times \sqrt{3}$ R30° Bi structure on the underlying Ir(111) surface. *Ab-initio* calculations show that this Bi structure is the most energetically favorable, and also illustrate that STM measurements are most sensitive to C atoms in close proximity to intercalated Bi atoms. Additionally, Bi intercalation induces a band gap ($E_g = 0.42$ eV) at the Dirac point of graphene and an overall *n*-doping (~0.39 eV), as seen in angular-resolved photoemission spectroscopy. We attribute the emergence of the band gap to the dislocation network which forms favorably along certain parts of the moiré structure induced by the graphene/Ir(111) interface.

Graphene as a two-dimensional material is of great interest due to its remarkable electronic and structural properties[1]. Its unique crystal structure includes an A-B sub-lattice symmetry that causes the charge carriers to act as massless Dirac fermions[2, 3], making it, along with a more general class of two dimensional materials, an interesting candidate for applications in nanoelectronics[1, 4]. Consequently, it is of technological importance to be able to manipulate the electronic properties of graphene, analogous to semiconductors. Two main goals of such manipulations are to induce a band gap at the Dirac point[5] leading to a semiconducting phase and ultimately to induce spin-splitting of the Dirac cone for spintronic applications[6].

There have been several approaches aimed at engineering a tunable band gap in graphene, which can be classified according to the inducing feature. In general, a band gap can be induced by the interaction with certain substrates (such as single crystal Cu(111) and Cu monolayers[7, 8]); by the structural confinement in graphene nanoribbons[9][10]; by the stacking of multiple graphene layers [11, 12]; by the periodic modulation of the graphene lattice (breaking the sub-lattice symmetry), which can be achieved by using patterned substrates[13], or through the patterned adsorption of other elements such as hydrogen[5].

Pristine graphene shows weak spin-orbit interactions[14], but increasing the strength of these interactions could give rise to new interesting possibilities in spin-based nanoelectronics. One route toward manipulation of graphene's electronic structure is intercalation of defined elements[15-17], between graphene layers weakly coupled to metallic surfaces. For example, it has been shown that intercalation of cobalt on graphene/Ir(111) leads to the existence of single layer magnetic Co islands and an induced spin polarization in the graphene layer[18]. First-principles calculations have predicted that intercalation of Bi in combination with strain can induce strong spin-splitting[19].

We report on the effect of Bi intercalation on the morphological and electronic structure of chemical vapor deposition (CVD) grown graphene on Ir(111) (G/Ir, in short). Bi deposition onto a clean G/Ir layer and subsequent annealing, as seen by scanning tunnelling microscopy (STM), results in an intercalated Bi layer in between graphene and the Ir(111) surface (G/Bi/Ir, in short). Structurally, the moiré superstructure produced by the G/Ir interface is still observed after Bi intercalation. Additionally, the Bi induces a periodic $\sqrt{3} \times \sqrt{3}$ R30° structure and a line dislocation network that tends to form along the bright areas of the moiré superstructure, seen in STM, which are located at the edges of the moiré unit cell. Using *ab-initio* calculations based on the Density Functional Theory (DFT), we show that the imaged morphology results from C atoms which are lying above Bi atoms in the intercalated layer, and that the intercalated structure is energetically favorable as compared to a surface alloy. Angle-resolved photoemission spectroscopy (ARPES) measurements show that Bi intercalation leads to an *n*-doping of the graphene as seen by a downward shift of the Dirac point energy ($E_{DP}$). Moreover, at $E_{DP}$, there is the emergence of a sizeable band gap $E_g = 0.42$ eV. Our calculations reproduce the *n*-doping and the observed band gap when considering a strained surface.

The Ir(111) surface was cleaned in ultra-high vacuum (UHV) by repeated cycles of Ar sputtering, followed by annealing to 1470 K. Additionally, carbon contamination was removed, when necessary, by intermittent annealing in $O_2$. A monolayer (ML) of high quality graphene was grown by exposing the clean Ir(111) surface to ethylene gas at a surface temperature of 1075 K. The growth process was followed by a short ($< 30\ s$) rapid heating to $T_{max} = 1455$ K. This step increases the quality of the resulting graphene ML. Bismuth was subsequently deposited on the G/Ir(111) surface at a sample temperature of 715 K. After the deposition, the sample was heated to a temperature of 1273 K for 60 s. This step leads to intercalated bismuth underneath the graphene. These surface regions where this intercalation structure is present are referred to as G/Bi/Ir throughout the text.

Scanning tunneling microscopy (STM) was performed using a home-built variable temperature STM in a UHV system with a base pressure below $1 \times 10^{-10}$ mbar [20,21]. Tip and sample were cooled to $T = 30$ K. Electrochemically etched and flashed W tips were used for all STM measurements. STM topography was recorded in constant current mode ($I_t$) with the bias applied to the sample ($V_S$). Differential conductance maps (short $dI/dV$ maps) were recorded by applying a modulation voltage ($V_{mod}$) using lock-in detection at a modulation frequency of $f = 5.477$ kHz.

ARPES data were acquired at the SGM3 beamline at ASTRID2[22]. The G/Bi/Ir samples were prepared and characterized within the STM chamber system and transported to the beamline in an ambient atmosphere for $\approx 12$ h. The samples were subsequently annealed to 575 K for 13 min in UHV before data acquisition. ARPES measurements of G/Bi/Ir were conducted at room temperature with photon energy of 47 eV. The pristine system, G/Ir, was measured for comparison at $\approx 85$ K using the same photon energy (47 eV).

Density functional theory (DFT) calculations were performed with the real-space projector augmented code wave GPAW [23]. The optB88-vdW approximation [24] to the exchange-correlation functional was used, which takes into account the dispersive interactions relevant for the investigated materials and has been successfully used to study other intercalated graphene systems [25]. A (4 × 4) supercell of graphene on a $\sqrt{12} \times \sqrt{12}$ R30° Ir(111) slab was used to explore the relative stability and electronic structure of G/Bi/Ir with varying Bi coverages. The graphene lattice constant was fixed to its optimized value of 0.1420 nm (atom-atom) 0.2465 nm (A-B sub-lattice constant), and the Ir lattice constant was adapted accordingly, resulting in strain of 2.76%. Such expansive strain increases surface reactivity following a concurrent shift of the metal d states towards higher energies[26]. Nevertheless, for Bi on Ir(111), the 2.76% expansive strain is expected to affect adsorption and intercalation energies in a similar way at different coverage, thereby not significantly affecting the conclusions of this work. Five Ir layers were used to describe the Ir(111) slab with the bottom iridium layer kept fixed, whereas the rest of the atoms of the system were allowed to relax until the force on each atom was lower than 0.2 eV/nm. A $(4 \times 4 \times 1)$ $k$-point mesh was used to sample the reciprocal space and a grid-spacing of 1.85 nm$^{-1}$ has been applied. In addition, a larger $(10 \times 10)$ graphene supercell on a $(9 \times 9)$ Ir(111) slab was used to model the moiré superstructure. 2D periodic boundary conditions were used parallel to the surface and a vacuum region of about 0.6 nm was included to separate the slabs from the cell boundaries. The Tersoff-Hamann approach [27, 28] was used to simulate the STM images of selected structures in the non-periodic direction perpendicular to the surface. For the models used, the band structure along the special $k$-vectors of the primitive cell of graphene was folded into the smaller Brillouin zone of the supercell. In order to recover the primitive cell picture of the band structure, the band unfolding technique proposed by Popesu and Zunger[29] was used as implemented in the BandUp code[30]. Since this implementation is only available for codes with plane-wave basis sets, the unfolding was performed using the charge density and wave functions obtained with single-point calculations using the VASP code[31, 32] and the same exchange-correlation functional.

Fig. 1(a-b) illustrates STM images of a full graphene mono layer on Ir(111) (G/Ir), which is identified by the characteristic moiré pattern[33], with the hexagonal moiré lattice with a lattice constant $\approx$ 2.5 nm, and atomic resolution of the honeycomb lattice of graphene. The quality and extensive coverage of the graphene layer, which nearly saturates the Ir(111) surface, was confirmed over several locations on the surface within a 1 mm² area. Upon low coverage Bi deposition and annealing, single layer islands of Bi can be seen underneath the graphene layer (Fig. 1(c)). The intercalation effect after heating deposited Bi, can be compared to the intercalation effects for other materials for example Co[18], Pb[34], and Eu[35] deposition on G/Ir(111), although with different microscopic structures.

The apparent height of the G/Bi/Ir islands is 0.192 nm. It is independent of the utilized $V_S$ and $I_t$, indicating that this apparent height is most likely topographic in origin. We note that this apparent height is too small to be a Bi bilayer which is typically favorable for Bi growth[36]. In addition to the same moiré structure as on G/Ir, G/Bi/Ir regions show a pronounced line dislocation network, which is absent on G/Ir (Fig. 1(c)). We discuss this dislocation network in more detail below. It is important to note, that the preservation of the moiré structure indicates that the Bi structure is commensurate with the Ir(111) surface lattice. We discuss the Bi structure in more detail below. By controlling the initial Bi coverage, we can vary the G/Bi/Ir content from single islands to nearly a full layer (Fig. 1(d)).

Upon closer inspection, G/Bi/Ir regions show two particular morphological characteristics in contrast to G/Ir (Fig. 2(a)). First, there is a hexagonal lattice, with a lattice constant of 0.47 nm as seen in fast Fourier transformation (short FFT) of the imaged G/Bi/Ir structure (Fig. 2(b)). By comparing this lattice to nearby G/Ir regions, we determine a 30° rotation with respect to the Ir(111) surface unit cell orientation. The lattice constant of the well-ordered hexagonal patches is related to the Ir(111) lattice constant by a factor of $\sqrt{3}$ (0.47 nm = $\sqrt{3} \times a_{\text{Ir}}$). As we demonstrate below through *ab-initio* calculations, STM constant-current imaging is more sensitive to the carbon atoms which reside directly on top of a Bi atom, allowing us to directly image the underlying Bi/Ir(111) interface. Each maximum within the hexagonal lattice indicates that the Bi produces a $\sqrt{3} \times \sqrt{3}$ R30° ($\sqrt{3}$) structure on the Ir(111) surface, similar to what is seen for G/Eu/Ir(111)[35, 37], excluding the dislocation network. We note here, within the investigated coverages and annealing temperatures, that we only see a $\sqrt{3}$ phase. In accordance with a $\sqrt{3}$ structure, we define a full layer of Bi, as 0.33 ML coverage of Bi. It is important to note that by imaging alone, we cannot distinguish between a $\sqrt{3}$ surface alloy where Bi periodically replaces surface Ir atoms, or a $\sqrt{3}$ Bi structure where Bi resides at hcp or fcc hollow sites of the Ir(111) surface lattice (Fig. 2(e-f)). Nevertheless, the *ab initio* calculations discussed below confirm that the latter case is thermodynamically favored.

In addition to the $\sqrt{3}$ hexagonal structure, the G/Bi/Ir regions show a three-fold symmetric dislocation network characterized by a meandering double line pattern, where the distance between lines is mostly ∼0.72 nm and the lines are imaged with a larger apparent height compared to the local hexagonal network. Bi covered regions can be easily distinguished from G/Ir regions by the $\sqrt{3}$ structure, the dislocation network, and by looking at the differential conductance ($dI/dV$) at particular voltages, as illustrated in Fig. 1(e-f). The dislocation network appears on all Bi regions, independent of the investigated coverage. As we image this dislocation network at all $V_S$ and $I_t$, we conclude the C atoms are topographically higher along the boundaries, i.e. along the dislocation lines, inducing a patterned strain network. As seen in FFT filtered images, the moiré pattern is observed in addition to the dislocation network (Fig. 2(c-d)). Upon closer inspection, the dislocation network traces out the higher boundaries of the moiré pattern. The orientation of a given double-line structure is rotated by 30° with respect to the Ir(111) lattice vectors. We note that the FFT taken from Fig. 2(a) shows there is a smearing of the intensity of the hexagonal lattice at larger *k*-values, which results from the combination of multiple domains, distortion of the $\sqrt{3}$ bismuth structure, and the dislocation network.

Within each $\sqrt{3}$ domain, the hexagonal structure exhibits few faults or variations as visible in Fig. 2(a). The hexagonal order is only disrupted when crossing a dislocation line, which results in a separation of $\sqrt{3}$ domains. The apparent atomic spacing is (a) 0.47 nm for atoms within a domain and (b) 0.72 nm for atoms in adjacent domains. It is important to note that shifting the adsorption site of Bi to the neighboring identical hollow site in the $\sqrt{3}$ structure ($a_{Ir}$ = 0.28 nm) results in a Bi-Bi spacing of 0.72 nm. This matches the observed atomic spacing at the dislocation network boundaries, suggesting that the dislocation lines separate adjacent fcc/fcc or hcp/hcp domains. While we cannot confirm the mechanism behind the dislocation network, it is most likely the result of the combination of the lattice mismatch of G to the $\sqrt{3}$ Bi/Ir interface, and the variation in bonding between the G regions which reside above Bi atoms vs. Ir atoms which produces local $\sqrt{3}$ domains.

In order to further understand the morphological and electronic structure of G/Bi/Ir, we performed DFT calculations, focusing on the structural characterization in comparison with STM. We first establish the favorable binding site of Bi with respect to the Ir(111) surface, and consider various orientations of the G unit cell with respect to the underlying interface. The Bi coverage of 0.08 ML, resulting from placing one Bi atom in the (4 × 4) graphene on $\sqrt{12} \times \sqrt{12}$ R30° Ir(111) supercell, has been used to explore the preferred intercalation sites of individual Bi atoms. In absence of the graphene ML, Bi preferentially adsorbs at hcp hollow sites, whereas the adsorption at fcc sites is 170 meV less stable. Under graphene, the stability of intercalated bismuth atoms also depends on the relative position of Bi with respect to the carbon atoms, but hcp sites are still more stable than fcc. Considering the various hcp sites with respect to the G structure, the most stable Bi intercalation site is the hcp site of Ir(111) which is located at the center of 4 carbon atoms. Sites where the Bi atom lies directly under a carbon atom, under a bridging position between two carbon atoms, or directly under the center of the graphene ring are $14 - 21$ meV less stable.

In order to assess the thermodynamic stability of different interface structures, we calculated the intercalation and alloying energies ($E_{int}$ and $E_{alloy}$) for G/Bi/Ir, defined as:

$$E_{int} = E(G/Bi_n/Ir) - E(G/Ir) - nE(Bi)$$

$$E_{alloy} = E(G/Bi_n/Ir_{m-n}) - E(G/Ir_m) - nE(Bi) + nE(Ir)$$

where $E(G/Bi_n/Ir)$ and $E(G/Bi_n/Ir_{m-n})$ correspond to the energy of the models with $n$ intercalated and alloyed Bi atoms, respectively, $E(Bi)$ is the energy of rhombohedral bulk Bi, and $E(Ir)$ is the energy of bulk Ir. Note that negative $E_{int}$ and $E_{alloy}$ values correspond to stabilizing interactions. Bi coverages of 0.08, 0.25, 0.33, 0.42, 0.50, and 0.66 ML, were considered. The Bi coverage is defined as the ratio between the number of Bi atoms and outermost Ir surface atoms. In order to determine the most stable distribution of Bi atoms for each coverage, we performed a global minimum search by means of an automated genetic algorithm[38]. The optimized structures obtained for intercalated Bi are shown in Fig. 3(a) and the alloying and intercalation energies of Bi are illustrated as a function of coverage in Fig. 3(b). Both the total adsorption energy per Bi atom (average) and the energy gain for each coverage increase (differential) are shown. At 0.08 ML, the formation of a surface alloy with Ir(111) is more stable than simply intercalating Bi between graphene and Ir(111), whereas for higher coverage the intercalated case is more stable. In addition, the most stable Bi coverage corresponds to 0.33 ML, which gives rise to a fully covered $\sqrt{3}$ pattern like the one observed in STM. The high stability of the 0.33 ML coverage indicates that Bi atoms will cluster at the G/Ir interface until the $\sqrt{3}$ structure is formed. The assignment of the G/Bi/Ir $\sqrt{3}$ structure with intercalated Bi is further confirmed by comparing calculated and measured heights. The presence of Bi increases the height of the graphene layer with respect to the G/Ir system. Although the resulting Bi-G distances are similar for both intercalated and alloyed Bi (~0.35 nm), the lower height of alloyed Bi (which is embedded in the Ir(111) surface) leads to lower G-Ir distance. As a result, for the 0.33 ML case, intercalated (alloyed) Bi leads to a difference in optimized height with respect to the G/Ir system of 0.22 nm (0.08 nm), which is in better (worse) agreement with the experimentally measured 0.19 nm. Therefore, on the basis of energetic and structural arguments, our calculations indicate that the preferred structure for the G/Bi/Ir system corresponds to intercalated Bi atoms occupying hcp positions arranged in a $\sqrt{3}$ structure. We also note that the calculated Bi-graphene distance is characteristic of physisorbed graphene[39, 40], which indicates that the interaction with Bi is rather weak. It is important to note that, only based on the experimental data, we cannot conclude if we observe fcc or hcp type domains.

In order to better compare it to the STM experiments, we optimized the moiré supercell model with intercalated Bi in a $\sqrt{3}$ configuration. This larger model consists of a (10 × 10) supercell of G on a (9 × 9) supercell of Ir(111), with Bi atoms occupying 1/3 of the hcp sites (0.33 ML). Fig. 3(c) illustrates the simulated STM image for the displayed model generated by means of the Tersoff-Hamann approach ($B_S = -0.2$ V, $I_t = 0.01$ nA). The position of the ordered Bi atoms is directly discernible in the form of large protrusions, giving rise to a pattern in perfect agreement with the experimental STM images. Thus, the presence of a Bi atom directly underneath the small protrusions of the C atoms increases the apparent height in STM images. This indicates that the interaction with Bi does indeed perturb the electronic structure of the graphene layer, and in the STM constant-current images the C atoms which are close to the intercalated Bi atoms are most visible (pronounced). This explains why the STM topography illustrates only a subset of C atoms of the graphene sheet. In simulations, an additional set of smaller but more intense protrusions corresponding to the C atoms of the graphene layer also appears, and their intensity depends on the relative position of each C atom with respect to the nearest Bi. We must note, however, that the simulated STM images are sensitive to the choice of current ($I_t$) and potential ($V_S$).

In order to characterize the changes in electronic structure induced by Bi intercalation, we performed ARPES measurements of the same samples illustrated above after STM characterization. Fig. 4 shows the ARPES data collected around the Dirac cone for G/Ir and for G/Bi/Ir. The comparison in the dispersion of the $\pi$-band of graphene before and after the Bi intercalation is shown in Fig. 4(b,d). The Dirac cone dispersion is shown along the high symmetry directions $\Gamma - K - M$ and $A - K - A'$ (schematics of the Brillouin zone in Fig.4(e)). Along the $\Gamma - K - M$ direction only the left branch of the cone is visible because of matrix elements effects[41], while the $A - K - A'$ direction is chosen for the isotropy in terms of intensity and band dispersion. In Fig. 4(a,c) the constant energy surface of the Dirac cone at 500 meV below the Dirac point energy, $E_{DP}$, is displayed for G/Ir and G/Bi/Ir respectively. For G/Ir, $E_{DP}$ is situated near $E_F$ and therefore the upper cone is not visible. For G/Bi/Ir, the upper part of the Dirac cone becomes clearly visible (Fig. 4(d)). This indicates that Bi intercalation has a sizeable $n$-doping effect on graphene. By fitting the $\pi$-band, we estimate a 0.390(3) eV down shift of the cone in respect to G/Ir.

In addition to the *n*-doping, Bi leads to a large band gap at $E_{\text{DP}}$. (Fig. 4(f)). A careful analysis close to the band edge reveals a deviation from the linear behavior as the upper and lower cones adopt a parabolic shape resulting from the opening of the energy gap. To further investigate this feature we find the location of the spectral shape maxima of the upper and lower cones by fitting the energy distribution curves (EDC) (symbols in Fig. 4(f)). The lower and higher branches are then fitted with a hyperbola to obtain the precise band shape (black solid line in Fig. 4(f)). By doing this, we determine that the virtual Dirac point, defined as the center point of gap, is at 0.390 eV below $E_{\text{F}}$ for G/Bi/Ir. The energy gap width, $\Delta E$, which is estimated from the distance between the two vertices of the hyperbolae, is $\Delta E \sim 420(20)$ meV. The overall broadening of the graphene spectral features compared to the G/Ir(111) situation is due to the surface lattice disorder introduced by intercalation and to the hybridization with the underlying Bi atoms. The intensity profile across the gap around the *K* point is shown in Fig. 4(g). This depicts clearly the presence of the double spectral feature, due to the band gap opening. The EDC across the energy gap can accurately be fitted by two Lorentzians multiplied by the Fermi function, without the need of introducing additional in-gap spectral features. The opening of such a large gap results in a calculated effective mass of $\pm 0.0527(2)m_e$, where $m_e$ is the free electron mass, at the *K* point of the graphene $\pi$-band.

For G/Ir, replica bands are found due to the moiré pattern (displayed in Fig. 1(a-b)) [42]. At the same time, mini gaps are created at the avoided crossing points between the Dirac cone and the replica bands. Furthermore, a small deviation in the linear dispersion is visible at $E_{\text{F}}$ because of the hybridization of the cone with the Ir(111) surface state[43] (Fig. 4(b)). In addition to electron doping and gap opening, Bi intercalation leads to the disappearance of these features: The Dirac cone dispersion is highly linear, no hybridization gaps are observed (Fig. 4(d)) and the replica bands disappear (Fig. 4(c)). This indicates that Bi intercalation decouples the graphene from the Ir(111) surface[44]. The interaction of the Bi atoms with the underlying Ir(111) surface furthermore leads to the suppression of the Ir(111) surface state.

To understand the origin of the changes of the electronic structure induced by Bi intercalation, we performed electronic structure calculations. Here, supercell models are necessary for describing the interaction between systems with mismatched lattices such as graphene and Ir and for investigating the different concentration (coverage) of defects or adsorbates such as Bi. However, as the supercell size increases, the corresponding reciprocal lattice shrinks. As a result, the bands of the reciprocal lattice of the primitive cell (first Brillouin zone) are folded in the smaller reciprocal lattice of the supercell. Such folded band structures are difficult to interpret and differ from their primitive cell counterparts and from measured ARPES spectra. The comparison to experiment and the search for band gaps has therefore traditionally been limited to an analysis of the density of states (DOS) rather than the actual band structure. Recently developed band unfolding techniques, however, allow recovering the primitive cell picture of the band structure[30] or the so-called Effective Band Structure (EBS), which can be directly compared to ARPES data. In order to further scrutinize the measured $n$-doping and band gap measured for the graphene/Bi/Ir system, we performed the band unfolding on a selected set of models including freestanding graphene and two cells with the intercalated $\sqrt{3}$ Bi structure: a simple supercell of G/Bi/Ir and the large supercell representing the moiré lattice. We note that higher intensities at a given $k$-vector of the primitive cell of the EBS indicate a larger number of bands crossing equivalent points of the reciprocal supercell.

Not surprisingly, the EBS along the $A - K - A'$ direction for non-distorted freestanding graphene shown in Fig. 5(a) perfectly reproduces the $\pi$-band dispersion and the Dirac point at the Fermi level. In turn, the band structure along the same direction calculated for the fully covered $\sqrt{3}$ G/Bi/Ir structure (corresponding to the 0.33 ML structure in Fig. 3(a) differs from freestanding graphene and reproduces the *n*-doping observed in ARPES (Fig. 4(d)). However, the calculated down shift of the bands (0.245 eV) is smaller than the measured one (0.390 eV) and the Dirac cone for this system is practically undistorted, with no indication of a band gap opening. The smaller down shift and the absence of a band gap in the EBS indicate that the interaction of the graphene layer with the underlying $\sqrt{3}$ Bi structure alone, and the resultant symmetry breaking, is not solely responsible for the observed band gap. This is in agreement with the calculated band structure for graphene on Bi(111)[30] and for Bi intercalated graphene on SiC(0001)[19], which also exhibit gap-less Dirac cones at *K*. We note that the unfolding for the models with higher coverage did not reveal a band gap opening either (not shown). Another possibility is that the band gap originates from the different periodicity of the moiré pattern, but the EBS for the larger supercell representing the moiré lattice also gives rise to an undistorted Dirac cone at *K* (not shown). We therefore discard higher Bi concentrations or the moiré pattern as responsible for the measured band gap, which leaves the dislocation network of the G/Bi/Ir system observed via STM as the most likely origin of the band structure distortion.

The dislocation network, which is only quasi-periodic and, therefore, incommensurate, would require supercells made of thousands of atoms to be modeled. Nevertheless, we hypothesize that the dislocated structure gives rise to distorted Bi-C distances that affect the band structure of graphene. In this line, strong spin-orbit couplings were predicted for Bi intercalated graphene on SiC(0001), but only when the graphene-Bi distances were contracted from their equilibrium values[19]. Here, we assess the effect of such a contraction on the band structure of the $\sqrt{3}$ G/Bi/Ir model by artificially displacing the graphene layer downwards 0.07 nm from its optimized equilibrium height. We note that the equilibrium Bi-G distance (0.33 nm) is typical of physisorbed graphene, which is dominated by vdW interactions. In turn, the Bi-G distances resulting from a 0.07 nm displacement are more characteristic of chemisorbed graphene, which is expected to hybridize more with the substrate, significantly altering the electronic structure of graphene and even leading to band gap fomation[39, 40, 45]. The optB88-vdW functional used in this study is able to properly describe these two bonding regimes in graphene-metal contacts[46] and remarkably, such a distortion has a relatively small energy cost (~80 meV per graphene unit cell). Nevertheless, the effect on the electronic structure is dramatic (Fig. 5(c)). Due to the shorter Bi-C distances of the chemisorbed state, graphene is more hybridized with Bi, broadening the $\pi$-band and further down shifting the Dirac cones to 370 meV with respect to the Fermi level. The linear dispersion of the $\pi$-band is also strongly distorted, leading to the appearance of a flat band that gives the lower part of the cone a pseudo-parabolic shape. Rather than a gap at *K*, this distortion opens small gaps near K both above and below the Dirac point. To calculate the full band structure including the strain network is beyond our computational capability. While artificially straining the lattice is not representative of our sample, it qualitatively illustrates that structural distortions leading to chemisorbed states, for example as produced by a dislocation network, can result in a band gap.

The investigation of bismuth-intercalated graphene on Ir(111) reveals a hexagonal $\sqrt{3} \times \sqrt{3}$ R30° surface structure rooted in an identical Bi structure at the Ir(111) interface [6, 34]. Bi intercalation simultaneously leads to a periodic dislocation network defined by a double atom wide line pattern which is quasi-three fold symmetric and traces out a particular region of the moiré pattern resulting from the G/Ir overlap. We confirm using DFT that the STM images C atoms which sit directly above Bi atoms on the Bi/Ir interface. Moreover, DFT shows that Bi favorably intercalates into a $\sqrt{3} \times \sqrt{3}\,R30°$ structure. Electronic structure characterization reveals that Bi intercalation leads to a strong *n*-doping of the graphene (~0.4 eV), along with a sizeable band gap at $E_{DP}$, where $\Delta E \sim 420$ meV. Moreover, ARPES data clearly illustrates that the band dispersion of G extracted from G/Bi/Ir, is more free-standing as characterized by a strong linear dispersion and an absence of replica bands, in comparison to G/Ir(111). Moreover, there is no clearly observed spin-splitting of the G bands, which suggests that spin-orbit coupling is not responsible for the band gap opening. Using DFT calculations, we observe a band gap opening only when distorting graphene-Bi contacts from their equilibrium distance. Therefore, we conclude that the strain network, and the structural distortions it leads to, are responsible for the opening of the band gap.

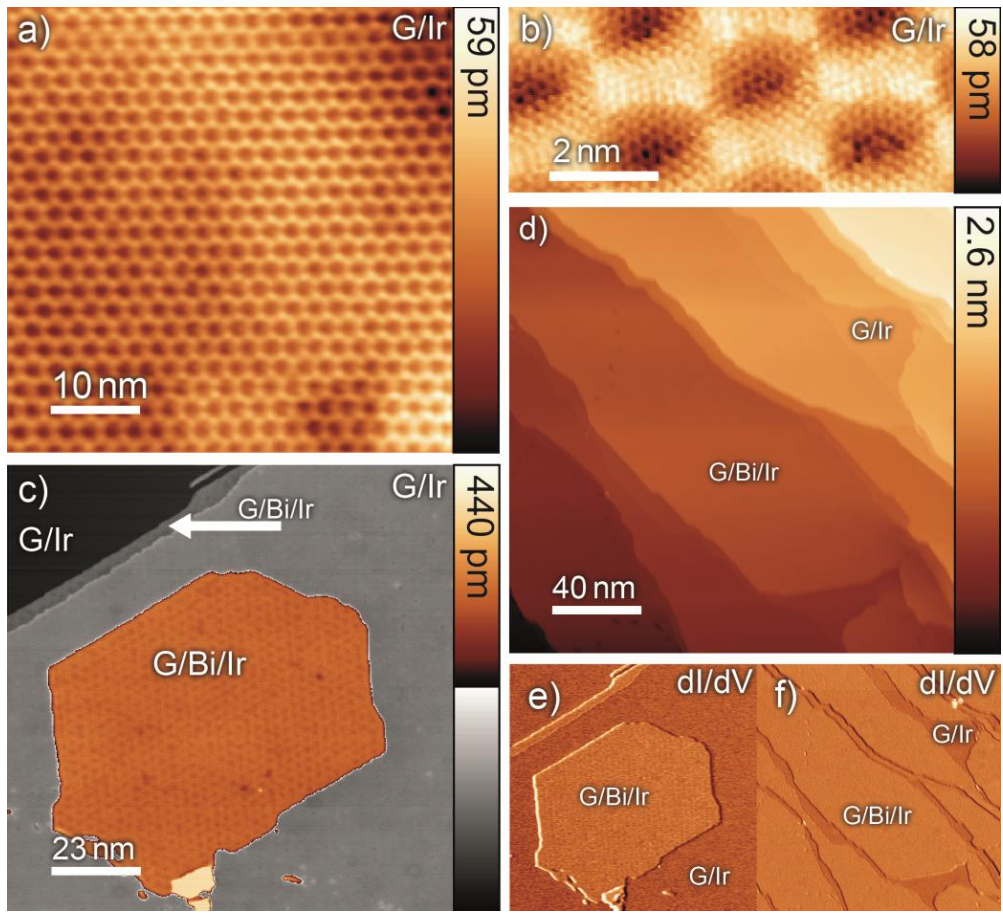

**Figure 1:** (Color online) Surface structure of the G/Ir and G/Bi/Ir system. STM constant current topographic images (a,b,c,d) and differential conductance maps (e,f) of graphene on Ir(111), (c,d,e,f) with intercalated bismuth of different coverage: (a) resolves the moiré superstructure of G/Ir after graphene growth (tunneling current $I_t = 35$ pA, sample bias voltage $V_S = 50$ mV). In addition (b) resolves the honeycomb lattice of the G/Ir system ($I_t = 290$ pA, $V_S = 480$ mV). (c) displays a G/Bi/Ir island and a thin G/Bi/Ir terrace (on the low side of the iridium step edge) on a G/Ir background; the G/Bi/Ir coverage on this sample is 0.05 ML ($I_t = 100$pA, $V_S = 500$ mV). (d) shows G/Bi/Ir terraces (on the low side of iridium step edges) on a G/Ir background; the G/Bi/Ir coverage on this sample is 0.95 ML ($I_t = 250$ pA, $V_S = 100$ mV). (e) and (f) display the contrast between areas of G/Bi/Ir to G/Ir in differential conductance $(dI/dV)$ map (modulation voltage $V_{mod} = 5$ mV for both), simultaneously acquired to the topographies (c) and (d).

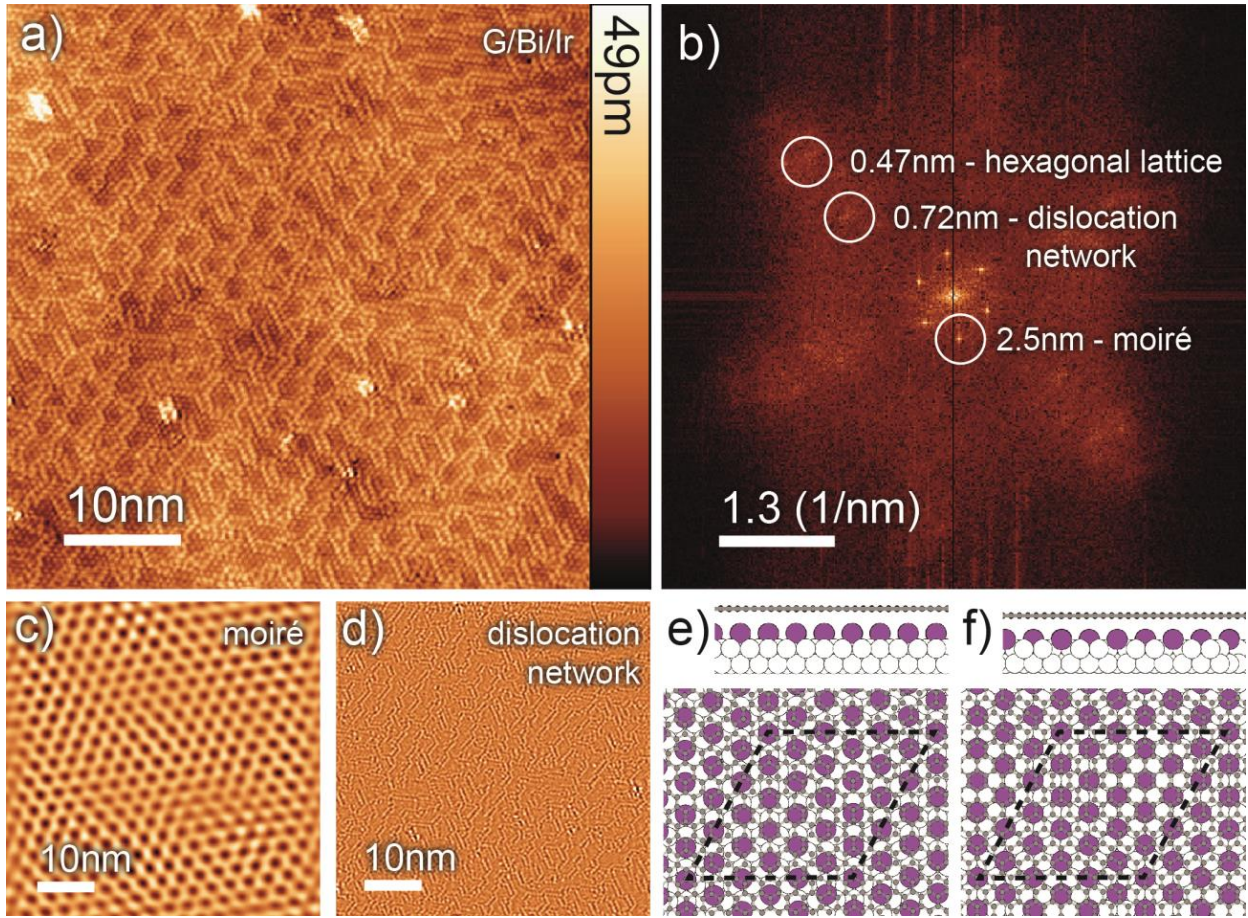

**Figure 2:** (Color online) Surface investigation of the G/Bi/Ir system. STM constant current topographic image (a) resolving the ($\sqrt{3} \times \sqrt{3}$) structure, the dislocation network and the moiré superstructure ($I_t = 684$ pA, $V_S = -10$ mV). Fast Fourier transformation (FFT) (b) of topography (a) featuring clear spots corresponding to the moiré and a halo corresponding to the ($\sqrt{3} \times \sqrt{3}$) structure and the dislocation network as marked[47]. (c) and (d) are FFT filtered versions of topography (a) restricted to the contribution of the moiré (c) and the dislocation network (d) (software WsXM 4.0[47]). Ball models (e) and (f) display the ($\sqrt{3} \times \sqrt{3}$) structure for a bismuth layer underneath the graphene for the intercalation (e) and surface alloy case (f).

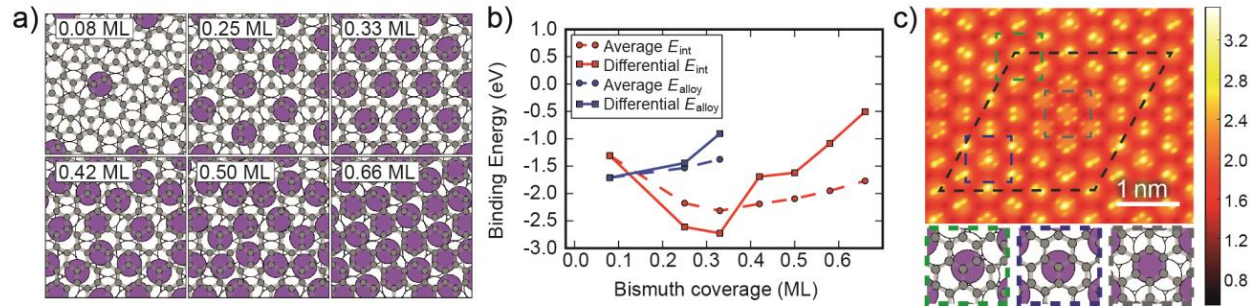

**Figure 3:** (a) Ball-stick models illustrating the most stable structures found for Bi (purple) adsorbed on Ir(111) (white) underneath graphene (grey) at different coverages. (b) Calculated average and differential $E_{int}$ and $E_{alloy}$ for increasing Bi coverages, revealing that systems with intercalated Bi are more stable than those where the Bi-Ir surface alloy is formed. The most stable system corresponds to 0.33 ML Bi coverage ordered in the ($\sqrt{3} \times \sqrt{3}$) configuration. Average and differential energies correspond to total adsorption energy per Bi atom and to the energy gain for each discrete coverage increase. (c) Tersoff-Hamann STM simulation ($V_S = -0.2$ V, $I_t = 0.01$ nA) for the moiré supercell model with a 0.33 ML coverage of intercalated Bi arranged in the ($\sqrt{3} \times \sqrt{3}$) structure. The bottom panels illustrate the local structure of selected regions, showing that C atoms in proximity of intercalated Bi have higher intensity. Color scale indicates corrugation with respect to average Bi height.

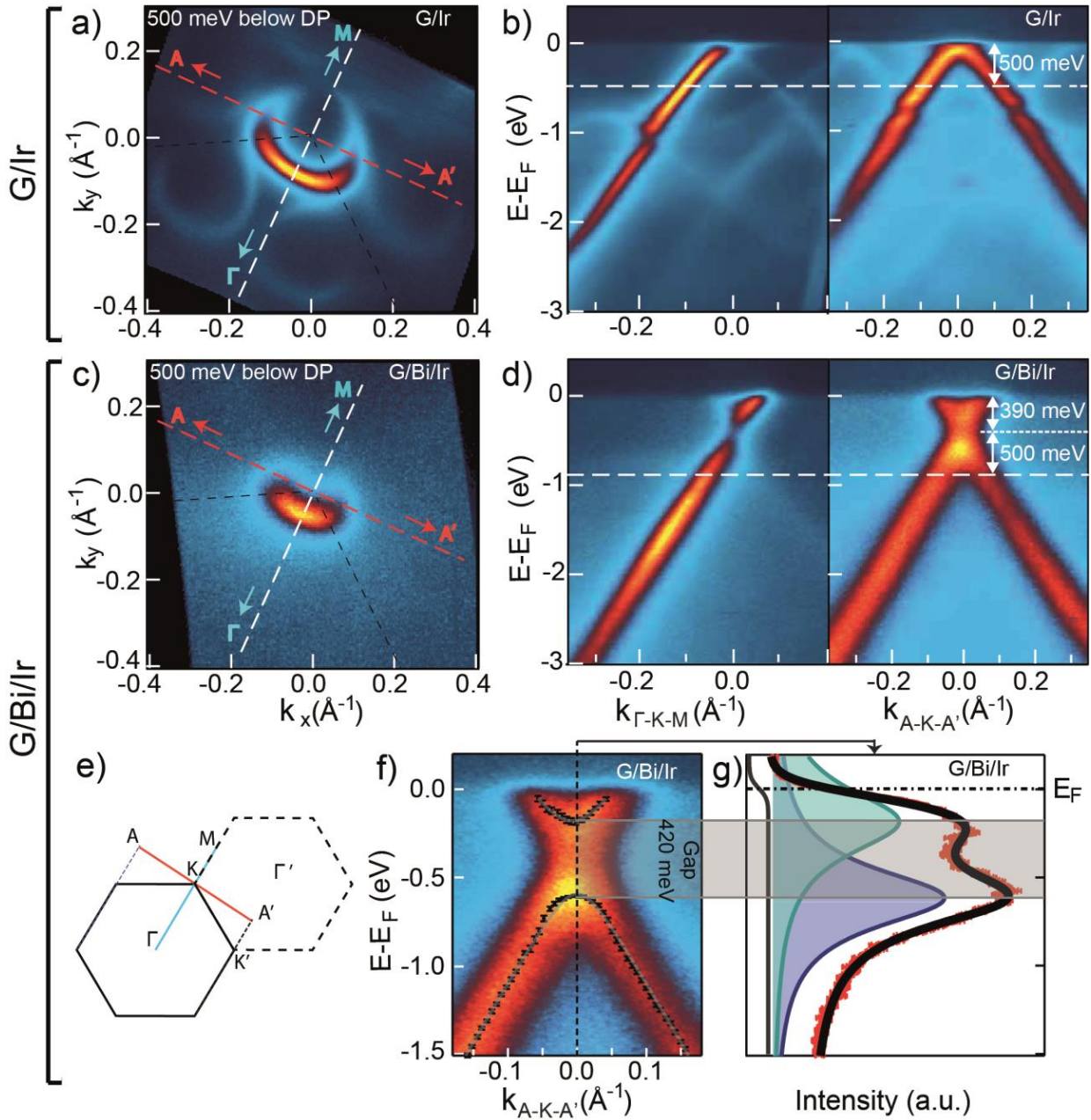

**Figure 4:** (Color online) Electronic structure around the *K* point of G/Ir (a,b) and of 0.95 ML G/Bi/Ir (c,d,f): (a)(c) Constant energy surface around the *K* point of graphene at 0.5 eV below the Dirac point (DP) position; black dashed lines indicate the edges of the first Brillouin zone and the colored dashed lines indicate the high symmetry directions along which the dispersion is shown. (b,d) Left panel: dispersion of the graphene Dirac cone along the *Γ-K-M* direction. Right

panel: dispersion of the graphene Dirac cone along the *A-K-A'* direction. The dashed line indicates the constant energy cut represented in (a) and (c). (e) Sketch of the graphene Brillouin zones with high symmetry points and directions. (f) Zoom-in into the Dirac cone in G/Bi/Ir close to the Fermi level along the *A-K-A'* direction. A 0.42 eV energy gap is opened at the level of the Dirac point. The symbols on top of the spectrum show the location of the energy distribution curves (EDC) maxima. The solid lines show the hyperbolic fitting on the upper and lower cones. (g) The red data points are the EDC intensity profile at the *K* point integrated over a 0.01 Å$^{-1}$ range and fitted by the black solid line, using two Lorentzians and a Fermi function cutoff. The two Lorentzian components of the fit are shown. The Fermi function is plotted on the left, an offset is applied for clarity. All the reciprocal space coordinates in the figure are rescaled with respect to the *K* point.

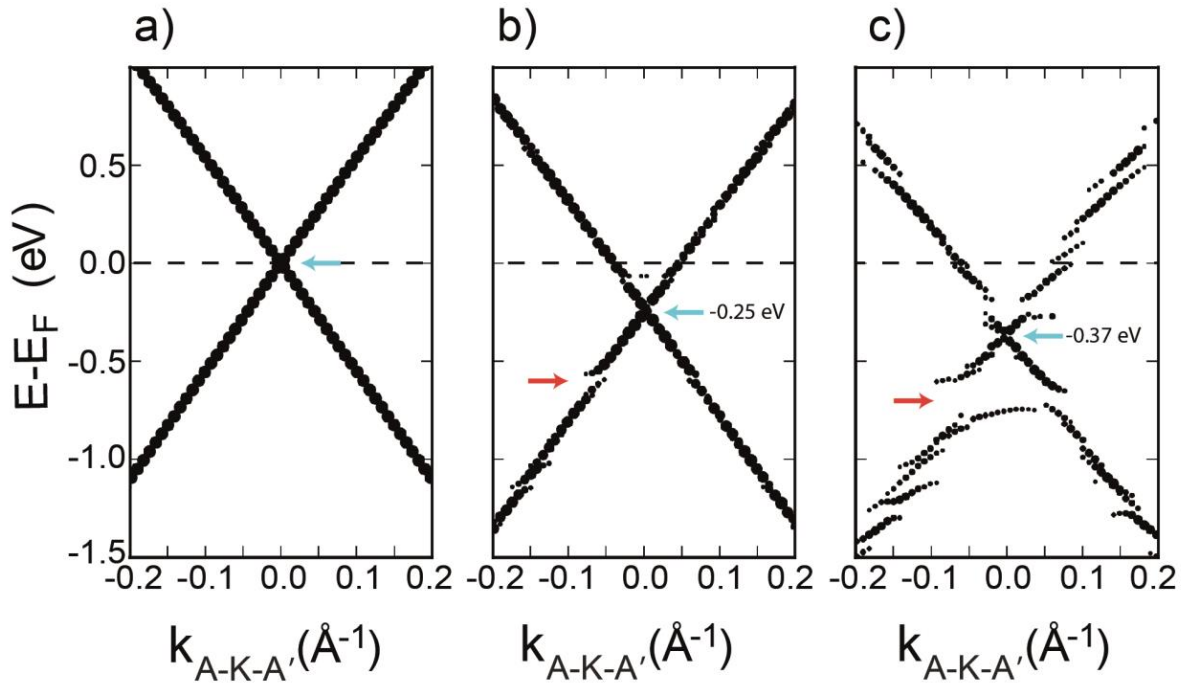

**Figure 5:** Unfolded band structure for (a) free-standing graphene, (b) Bi intercalated graphene with a $\sqrt{3} \times \sqrt{3}$ R30° interface structure (0.33 ML), and (c) for the same $\sqrt{3} \times \sqrt{3}$ R30° structure

under an artificial compressive strain applied in the surface normal direction. The size of each data point indicates the intensity (spectral weight) at each corresponding *k*-point and energy interval. In order to visualize the π-band more clearly, only data points with intensities above a given threshold are shown. Red arrows point to the induced gaps in the π-band and blue arrows to the center of the Dirac point. We note that the reduction from the equilibrium Bi-graphene distance of the stable physisorbed regime (b) to a value more characteristic of chemisorbed graphene (c) leads to strong distortions of the band structure of graphene, which is further n-doped and with the presence of gaps in the π-band.

**Author Contributions – from template**

JoW, MM, TH, and MB performed the experiments and analyzed the data. AB and BH performed the calculations. JoW, PH, and AAK designed the experiments. All authors helped write the paper.

**Funding Sources**

**JoW and AAK acknowledge funding from the Emmy Noether Program of the DFG (KH324/1-1) and the Vidi program from NWO. JW, PH and AAK acknowledge funding from the SPP1666 of the DFG. AB acknowledges support from the European Research Council under the European Union's Seventh Framework Programme (FP/2007-2013) / Marie Curie Actions / Grant no. 626764 (Nano-DeSign). PH also acknowledges the Danish Council for Independent Research (Sapere Aude program, Grant no. DFF-4002-00029) and the VILLUM foundation. T.H. acknowledges funding from the project HA 6037/2-1 of the DFG.**

ACKNOWLEDGMENT

JW and AAK thank Maciej Bazarnik, Andreas Sonntag and Andreas Eich for their input. AB is grateful to Dr. Paulo V. C. Medeiros for the open-source distribution of the BandUp code and for the useful technical support.

ABBREVIATIONS

STM, scanning tunneling microscopy; ARPES, angle-resolved photoemission spectroscopy; DFT, density functional theory; CVD, chemical vapor deposition; UHV, ultra-high vacuum; ML, mono-layer; Ir(111), iridium (111) surface; G/Ir, graphene on iridium; G/Bi/Ir, graphene on intercalated bismuth on iridium; FFT, fast Fourier transformation; $dI/dV$ map, differential conductance maps;